\documentclass[prl,twocolumn,showpacs,preprintnumbers,amsmath,amssymb]{revtex4}
\usepackage{amsfonts}
\usepackage{graphicx}
\usepackage{dcolumn}
\usepackage{float}
\usepackage{bm}

\newcommand{\com}[1]{}

\begin{document}

\title{Coherent phase slips in superconducting nanorings}

\author{D. Mozyrsky}\email{mozyrsky@lanl.gov}
\author{D. Solenov}

\affiliation{Theoretical Division (T-4) and the Center for
Nonlinear Studies (CNLS), Los Alamos National Laboratory, Los
Alamos, NM 87545, USA}

\date{\today}

\begin{abstract}

We study quantum fluctuations of persistent current in a small superconducting ring. Based on a microscopic model
of the ring we argue that under certain conditions such ring will exhibit coherent quantum phase slips, similar
to those in a flux qubit. We evaluate the frequency of such coherent oscillations and find that it is strongly
dependent on wire's diameter primarily due to the large momentum released by the condensate as a result of a
phase slip event. We also find that the value of such frequency is not a self-averaging quantity, that is,
it depends on a particular realization of the static impurity potential.
\end{abstract}

\maketitle

Relaxation of persistent currents in quasi one-dimensional (1D)
superfluids due to phase slip transitions has been a subject of
extensive studies over several decades. While at finite
temperatures phase slips are known to occur via thermal activation
through an effective barrier \cite{amb}, at sufficiently low
temperatures they are believed to be generated by the quantum
fluctuations of the order parameter, i.e., macroscopic quantum
tunneling. Some experimental studies of electric transport in
superconducting \emph{wires} at low temperatures have, indeed,
demonstrated finite residual resistivity in sufficiently thin
wires \cite{bez,atom}, consistent with the quantum phase slip
(QPS) picture \cite{Zaikin}. Other experimental studies \cite{exp},
however, did not confirm these results, suggesting,
in particular, that the experiments in Ref. \cite{bez} could be
explained in terms of thermally activated phase slips in the
presence of Coulomb blockade.

It has been argued in the literature \cite{mooij,naz} that the controversy
(of whether the observed phase slips are of quantum origin) can be
clarified by observing the coherent effects in fluctuations of
electric currents, - the thermally activated phase slips, obviously,
are not supposed to exhibit any phase coherence.
In particular, based on \emph{phenomenological}
arguments, it was proposed that short and thin superconducting
filaments can possibly operate as sources of \emph{coherent} QPSs
\cite{mooij,naz,buch}, just like Josephson junctions in flux
qubits. The experimental evidence for coherent QPS in Josephson
junction arrays has recently been reported in Ref. \cite{devore}.
Yet, no \emph{microscopic} theory of coherent QPS-driven current
fluctuations has been developed so far.

In this paper we provide a microscopic theory of coherent QPS
in a small superconducting (nano) ring. We argue that if a half-quantum of external
magnetic flux is applied to the ring, it will exhibit
coherent switching between states with different persistent currents, i.e.,
coherent QPS transitions. The
rate (frequency) of such QPS is controlled by the strength of the
impurity potential: For a translationally invariant system, i.e.,
in the ``clean'' limit, the QPS are suppressed due to the Landau
criterion \cite{prokof,khleb}. Indeed, a QPS event is accompanied
by the production of sufficiently large momentum, which (at zero
temperature) cannot be absorbed by the quasiparticles unless the
current (or superfluid velocity) exceeds its critical value. In
the presence of impurities the momentum conservation is
``violated'' and so the QPS can occur. However, as we show in this
paper, the rate of the QPS is proportional to the Born's
scattering cross-section of the condensate at the impurities,
which is small for large QPS momenta, i.e., for wires whose
diameter significantly exceeds the interparticle distance.
Moreover, the value of the QPS frequency turns out to be strongly
sample dependent: the ratio of its root-mean-square deviations to
its average value is a constant which does not decrease with
system's size, as in the case of incoherent QPS in sufficiently
long wires \cite{prokof,khleb2}.

Before proceeding to the evaluation of the QPS frequency for a BCS superconductor, let us evaluate this quantity in the framework of the Bose-Luttinger liquid. The calculation for this model is rather straightforward and serves as a reference point for the more complex superconducting case. The Hamiltonian for such quantum liquid confined to a 1D ring with finite circumference $L$ can be written as
$H=H_0+H_1$ \cite{haldane}, where
\begin{equation}\label{eq:linham}
H_0 = H_Q+H_J=\sum_{k\neq 0} \omega_k a^\dag_k a_k +(E_J/2)({\hat J}-J_0)^2
\end{equation}
is the Hamiltonian of the liquid in the absence of impurity
potential, i.e., for a translationally invariant system,  and
$H_1$ is due to the impurity potential. In Eq. (\ref{eq:linham})
$\omega_k$ is energy of a quasiparticle with momentum $2\pi k/L$
and $a^\dag_k$ ($a_k$) are bosonic operators that create
(annihilate) the quasiparticles. The $H_J$ term in
Eq.~(\ref{eq:linham}) is the energy associated with the superfluid
motion of the liquid and is characterized in terms of winding
numbers $J=0, \pm 1, ...$, i.e., eigenvalues of operator ${\hat
J}$. Constant $E_J$ is related to the effective inductance of the
liquid (typically of kinetic origin for both neutral systems and
ultrathin superconducting wires) and $J_0$ is the ``external''
winding number supplied by the magnetic field (for charged
particles) or by the rotation (for neutral particles). Note that
in Eq. (\ref{eq:linham}) the states with different $J$'s are
decoupled from each other owing to the Landau criterion.

The presence of $H_1=\int_0^L dx V(x)n(x)$, where $V(x)$ is
impurity potential and $n$ is 1D particle density, introduces a
matrix element between states with different $J$'s. This can be
seen by writing $n$ in terms of the displacement
phase $\Theta$ \cite{haldane} according to the relation
$n=n_0+\delta n$,
\begin{eqnarray}\label{eq:rho}
\delta n \sim \sum_{l>0} ({\hat J^-})^l \exp (il\Theta)+({\hat J^+})^l \exp (-il\Theta),
\\\nonumber
\Theta = 2\pi n_0 x + \sum_{k\neq 0}\left(2\pi c\alpha \over L\omega_k\right)^{1/2}(a_k^\dag + a_{-k})e^{2\pi i k x/L} ,
\end{eqnarray}
where operators ${\hat J}^{\pm}$ raise/lower the winding numbers
$J$ by $\pm 1$ , and $\alpha$ is the Tonks parameter,
$\alpha=\pi n_0/(mc)$, $m$ is particle's mass and $c$ is sound
velocity. In this paper we set both the Plank constant and the speed of
light equal to $1$. The value of $\langle 0|H_1|1\rangle$, where $|0\rangle$
and $|1\rangle$ are the states of the liquid with zero and one
winding numbers and zero quasiparticles, can now be easily
evaluated with the use of Eq. (\ref{eq:rho}). Retaining $l=1$
terms only in Eq. (\ref{eq:rho}), we obtain
\begin{equation}\label{eq:corel}
\langle 0|H_1|1\rangle\sim \left(\int_0^L dx V(x)e^{2\pi i n_0 x}\right) e^{- {\pi c \alpha\over L}\sum_{k\neq 0}{1\over\omega_k}}.
\end{equation}
If the value of this matrix element is small
compared to the gap between different quasiparticle states
$\Delta\omega=2\pi c/L$, the system's dynamics at low enough
temperatures (i.e., for $T\ll \Delta\omega$) reduces to the
transitions between $J=0$ and $J=1$ ``ground'' states (here and in
the following we are interested in the degenerate $J_0=1/2$ case,
which can always be achieved by the appropriate choice of magnetic
field or rotation frequency). That is, if the system is initially
set in a state with, say, $J=0$, its probability to occupy that
state, $|\langle 0(t)| 0(0) \rangle|^2$, will exhibit coherent
oscillations with frequency $\omega_0 = |\langle 0|H_1|1\rangle|$,
as in the case of a two-state system. Assuming the Bogolubov's
dispersion relation $\omega_k =c[(2\pi k/L)^2 + (2\pi
k/L)^4\xi^2/4] ^{1/2}$, where $\xi=(mc)^{-1}$ is coherence length,
we obtain that $\sum_n\omega_n^{-1}\simeq\ln{(2e^{1/2}L/\pi\xi)}$
in Eq. (\ref{eq:corel}) and
\begin{equation}\label{eq:omega}
\omega_0 \sim |V_{n_0}|\times(\pi\xi/2e^{1/2}L)^\alpha,
\end{equation}
where $V_{n_0}=\int dx V(x)e^{2\pi i n_0 x}$.
It is interesting that $\omega_0$ does not self-average with the
growth of $L$. Indeed, if we assume that the disorder potential is
short range and Gaussian with $\langle
V(x)V(x^\prime)\rangle=V_0^2\delta(x-x^\prime)$, we see that the
average frequency $\langle\omega_0\rangle = (\pi^{1/2}/2)n_0 V_0
L^{1/2} (\pi\xi/2e^{1/2}L)^\alpha$, while its root-mean-square
deviations  $(\langle\Delta\omega_0^2\rangle)^{1/2} \sim
\langle\omega_0\rangle$.  Therefore we predict that the frequency
of coherent oscillations is strongly sample-dependent. While this
fact may seem surprising, it is a direct consequence of systems's
coherence: Indeed, for a sufficiently large system, e.g., Refs.
\cite{prokof, khleb2}, phase slips occur at different,
disconnected from each other regions, which effectively leads to
self-averaging. In our case the phase slips are strongly
overlapping (note that $\omega_0\ll c/L$) and thus averaging over
the disorder does not occur.

Let us now consider the fermionic model. The Hamiltonian density for electrons within the standard BCS model can be written as
\begin{eqnarray}\label{eq:hamiltonBCS}
H ={1\over 2m}\psi^\dag_\sigma (i\nabla-e{\bf A}_0)^2\psi_\sigma + V({\bf r})\psi^\dag_\sigma\psi_\sigma\ \ \ \ \ \ \ \ \ \\\nonumber
+e^2\int d^3{\bf r}{\psi^\dag_\sigma({\bf r})\psi_\sigma({\bf r})\psi^\dag_{\sigma^\prime}({\bf r}^\prime)\psi_{\sigma^\prime}({\bf r}^\prime)\over |{\bf r}-{\bf r}^\prime|}+(\Delta\psi_\uparrow\psi_\downarrow+{\rm h.c.}),
\end{eqnarray}
where repeated index $\sigma$ means summation over $\uparrow$ and $\downarrow$ spin components of fermionic field $\psi_\sigma({\bf r})$. In
Eq. (\ref{eq:hamiltonBCS}) ${\bf A}_0$ is vector potential due to an external magnetic field and $V({\bf r})$ is impurity potential, to be specified below. The electrons are restricted to move in a wire of cross section $S_0=\pi r_0^2$ ($r_0$ is assumed to be much smaller than the superconducting coherence length $\xi_s$), which has the shape of a ring (or torus) with circumference length $L\gg \xi_s$. We assume that the magnitude of the superconducting order parameter $\Delta({\bf r})=|\Delta_0|e^{i\theta({\bf r})}$ is fixed and allow for variations of its ``soft'' phase $\theta$ only. Such assumption is justified by the existence of the gapless (i.e., sound-like) mode arising in a quasi-1D situation (see below). In Eq. (\ref{eq:hamiltonBCS}) we neglect the geometric inductance of the loop, which is small compared to the kinetic inductance for ultrathin wires.

Unfortunately it does not seem possible to directly introduce the
displacement phase $\Theta$ for the Cooper pairs as it was done in
the previous model. For the conventional phase $\theta$, however,
one can derive an effective low energy action within the
perturbative approach, i.e., assuming that gradients of $\theta$
as well as its time derivatives are small. In order to do so we
apply a gauge transformation $\psi_\sigma\rightarrow\psi_\sigma
e^{i\theta/2}$ to the Hamiltonian in Eq. (\ref{eq:hamiltonBCS})
and expand the corresponding action in terms of
$\partial_\tau\theta$ and $\nabla\theta$.

The first order contribution to the effective action for the phase
$\theta $ is the Berry phase term \cite{com}
\begin{equation}\label{eq:S1}
{\cal S}_1 = i\int d\tau d^3{\bf r}\langle \psi^\dag_\sigma({\bf
r})  \psi_\sigma({\bf r})\rangle  \partial_\tau\theta .
\end{equation}
In Eq. (\ref{eq:S1}) the averaging is taken with respect to the
ground state of the electrons, but {\it not} with respect to the
disorder potential $V$. The importance of particle density
variations has been discussed in Refs. \cite{prokof, khleb}, where
it was shown that the QPS processes are suppressed by the Berry
phase, e.g., Eq. (\ref{eq:S1}), in translationally invariant
systems. Such suppression is, essentially, a consequence of Landau
criterion, discussed above.
Note that the quantity $\int d\tau \partial_\tau\theta$ is a
topological invariant: It is determined by a discrete set of points
corresponding to the positions of the QPS rather than by the
particular dependence of $\theta$ on $\tau$ and ${\bf r}$; see
below.

The second order terms in $\partial_\tau\theta$ and $\nabla\theta$ have been derived in Refs. \cite {Zaikin}. For consistency we briefly outline the calculation. First one introduces the scalar potential $\phi$ by decoupling the Coulomb interaction term in Eq. (\ref{eq:hamiltonBCS}) via Hubbard-Stratonovich transformation. The second order terms obtained after the averaging over the $\psi$ fields are
\begin{eqnarray}\label{eq:S2}
{\cal S}_2 = \int d\tau d^3{\bf r} \left[\nu(i\partial_\tau\theta+2e\phi)^2 \right.\ \ \ \ \ \ \ \ \ \ \ \ \ \ \\\nonumber
\left.+ (\rho_s/2m)(\nabla\theta +2e{\bf A}_0)^2 + (\nabla\phi)^2/8\pi\right],
\end{eqnarray}
where $\nu$ is free electron density of states and ${\rho}_s$ is the density of superconducting electrons. Note that Eq. (\ref{eq:S2}) is
also averaged with respect to the impurity potential. Such averaging is valid since $\theta$ is a slowly varying quantity. Moreover, since the thickness of the wire is small compared to the superconducting coherence length, $\theta$ is effectively a 1D field. The field $\phi$ can now be excluded from Eq. (\ref{eq:S2}). In doing so one should keep in mind that $\phi$ is also nonzero in the region outside the wire. The contribution of the outside region can be shown to dominate over the inside region and the last term in Eq. (\ref{eq:S2}) can be written as $\int dx C\phi^2/2$, where $C\simeq \epsilon[2\ln{(L/r_0)}]^{-1}$ is the capacitance (per unit length) of the ring, $\epsilon$ is dielectric constant of the surrounding medium and the scalar potential $\phi$ inside the wire is assumed to be constant along the wire's cross section. Then $\phi$ can be integrated out and from Eqs. (\ref{eq:S1}, \ref{eq:S2}) we obtain an effective Lagrangian density for $\theta(\tau, x)$:
\begin{equation}\label{eq:L}
{\cal L} = in(x){\partial_\tau\theta} + {C\over 2e^2}(\partial_\tau\theta)^2+{\rho_s S_0\over 2 m}(\partial_x\theta +2eA_0)^2.\ \
\end{equation}
In Eq. (\ref{eq:L}) we have defined the 1D density $n(x)=\int d^2{\bf r}_\perp \rho({\bf r})$, where $\rho({\bf r})=\langle\psi^\dag_\sigma ({\bf r}) \psi_\sigma({\bf r})\rangle$, and assumed that $e^2\nu S_0 \gg C$. $A_0$ in the last term in Eq. (\ref{eq:L}) is the component of the external vector potential ${\bf A}_0$ along the direction of the wire.  The last two terms in Eq. (\ref{eq:L}) describe the Mooij-Schon gapless plasmon mode propagating with phase velocity $c_s=(\rho_s S_0e^2/mC)^{1/2}$ \cite{Zaikin}.

We are interested in the probability amplitude for the system to
remain in state $|0\rangle$, i.e., with zero current. If a half of
magnetic flux, $\Phi_0=\int dS B_0=\oint d{\bf x}{\bf A}_0$, is
applied, i.e., $\Phi_0\to \pi/2e$, the zero current state is
degenerate with the state with one winding number, $|1\rangle$,
i.e., with electric current equal to $2\pi e\rho_s/mL$. The
amplitude $\langle 0(\tau)|0(0)\rangle$ can be evaluated using
instanton method. The leading contribution is
\begin{eqnarray}\label{eq:prob}
\langle 0(\tau)|0(0)\rangle \sim 1 + K \int_0^\tau d\tau_1 \int_0^{\tau_1} d\tau_2\ \ \ \\\nonumber
\times\int_0^L dx_1\int_0^L dx_2 e^{-{\cal S}_{\rm cl} (\tau_1,x_1; \tau_2,x_2)}+...,
\end{eqnarray}
where action ${\cal S}_{\rm cl}$ is evaluated along the classical
trajectory, which starts in state $|0\rangle$, then, at time
$\tau_1$ passes on to the state $|1\rangle$, and at time $\tau_2$
returns to $|0\rangle$. The coefficient $K$ is related to the
determinant for the fluctuations around the classical trajectory.
It is independent of the first, topological term in Eq.
(\ref{eq:L}), and was estimated in Refs. \cite{khleb}: $K\sim
\alpha_s^2/(\xi_s^2\tau_0^2)$, where $\tau_0=\xi_s/c_s$ and
$\alpha_s = \pi (\rho_s S_0 C/e^2 m)^{1/2}$.

The classical trajectory action in Eq. (\ref{eq:prob}) corresponds
to a superposition of two phase slips (i.e., vortex at $\tau_1,
x_1$ and anti-vortex at $\tau_2, x_2$) \cite{Zaikin}. For a finite
ring we should account for the boundary conditions: The system is
periodic in $x$ direction (with period $L$) and infinite along the
$\tau$ direction (again we consider a zero temperature case). The
(anti)vortex solutions for the phase $\theta$ satisfying such
boundary conditions has the form
\begin{eqnarray}\label{eq:vortex-theta}
\theta(x,\tau) = \pm\mathrm{sgn}(\tau-\tau_{1(2)})\ \ \ \ \ \ \ \ \ \  \ \ \ \ \ \ \ \ \ \ \ \ \ \ \ \ \ \ \ \\\nonumber
\times\tan^{-1} \frac{ e^{-2\pi
c_s|\tau-\tau_{1(2)}|/L}\sin [2\pi (x-x_{1(2)})/L] }{ 1 - e^{-2\pi c_s|\tau-\tau_{1(2)}|/L}\cos [2\pi (x-x_{1(2)})/L] }
\end{eqnarray}
Note that in the limit $L\to\infty$ we recover the well known
expression $\theta(x,\tau) \to \pm \tan^{-1}[(x-x_1)/c_s(\tau-\tau_1)]$, e. g., Refs. \cite {Zaikin}.  Using
this solution and Eq.~(\ref{eq:L}) we evaluate ${\cal S}_{\rm cl}$
in Eq. (\ref{eq:prob}). After some calculation we find
\begin{eqnarray}\label{eq:act}
{\cal S}_{\rm cl} = 2\pi i\int_{x_2}^{x_1}dx\, n(x) +\eta\Delta\Phi\Delta\tau+ 2\alpha_s \ln{(L/\xi_s)}\ \ \ \\\nonumber
- 2\alpha_s \ln{\left(1+e^{-{4\pi c_s|\Delta\tau|\over L}}- 2e^{-{2\pi c_s|\Delta\tau|\over L}}\cos{2\pi\Delta x\over L}\right)},
\end{eqnarray}
where $\Delta\tau=\tau_2-\tau_1$, $\Delta x=x_2-x_1$, $\eta=4\pi e\rho_s S_0/mL$ and $\Delta\Phi = \Phi_0-\pi/2e$. The first term in Eq. (\ref{eq:act}) is due to the Berry phase in Eqs. (\ref{eq:S1}, \ref{eq:L}); it is controlled by the relative position of the vortices, but not by their internal structure. The second term in Eq. (\ref{eq:act}) is proportional to the bias energy $\eta\Delta\Phi$, i.e., the energy difference between states with and without current.  The last two terms are the the vortex self-energy and interaction energy respectively.

At the point of degeneracy (for $\Delta\Phi=0$), the interaction energy between vortices vanishes when $\Delta\tau = \infty$. Then the last term in the rhs of Eq. (\ref{eq:prob}) is equal to $\omega_0^2\tau^2/2$, where
\begin{equation}\label{eq:K}
\omega_0 = K^{1/2}  e^{-\alpha_s \ln{(L/\xi_s)}}|\int_0^L dx^\prime e^{2\pi i\int_0^{x^\prime}dx\,n(x)} |.
\end{equation}
Moreover, it is straightforward to verify that the next, four-vortex contribution in Eq. (\ref{eq:prob}) is $\sim (\omega_0\tau)^4/4!$, etc., and therefore $\langle 0(t)|0(0)\rangle=\cos(\omega_0 t)$, i.e., the system exhibits coherent oscillations with frequency $\omega_0$.

For a translationally invariant system (with $V=0$), $n(x)={\rm const}$ and therefore the integral over $x^\prime$ in Eq. (\ref{eq:K}) vanishes and $\omega_0=0$, as expected. For not too strong disorder and not too thin wires we may write $n(x)=n_0+\delta n(x)$, with $n_0\gg\delta n$. Then
\begin{equation}
\int_0^L dx^\prime e^{2\pi i\left[n_0 x^\prime + \int_0^{x^\prime}dx \delta n(x)\right]} \simeq {1\over 2\pi n_0}\int d^3{\bf r} e^{2\pi i n_0 x}\,\delta \rho({\bf r}),\nonumber
\end{equation}
where $\rho({\bf r})=\rho_0+\delta\rho({\bf r})$. Since
$\delta\rho({\bf r})\simeq\int d^3{\bf r}^\prime \chi({\bf r}-{\bf
r}^\prime)V({\bf r}^\prime)$, where $\chi$ is the static
electronic susceptibility, Eq. (\ref{eq:K}) is basically
equivalent to that for the bosonic model, e.g., Eqs.
(\ref{eq:corel}, \ref{eq:omega}). Note that the vortex
self-interaction term $-\alpha_s \ln(L/\xi_s)$ in Eqs.
(\ref{eq:act}, \ref{eq:K}) coincides with the last exponent
($\sim\sum_n \omega_n^{-1}$) in Eq. (\ref{eq:corel}), arising from
the evaluation of $\langle\exp{(i\Theta)}\rangle$ in the bosonic
problem.

Again, as in the bosonic problem, we expect that the QPS frequency
is strongly sample-dependent, i.e., it is a random number whose
average is comparable to its root-mean-square deviations. In order
to estimate the typical (average) frequency of QPS oscillations,
we must evaluate $\langle|\int dx e^{2\pi i n_0 x}\,\delta
n(x)|^2\rangle$. We assume that the disorder is due to random
impurities carrying charge $e$, i.e., $V({\bf r})=\sum_{i=1}^{N_i}
e^2/|{\bf r}-{\bf r}_i|$. The response function $\chi$ for
the Hamiltonian in Eq. (\ref{eq:hamiltonBCS}) is generally not
well known at finite wavevectors. However, the momentum $2\pi
n_0=2\pi \rho_0 S_0$ transferred from the condensate to the
impurities as a result of a QPS event is rather high compared to
both the Fermi wavevector $k_F$ and the inverse screening radius.
At such high momenta the electronic response function is that of
free electrons, $\chi(q)\simeq\chi_0^{q\rightarrow\infty}(q)
\simeq - 4m k_F^3/(3\pi^2 q^2)$, and therefore
\begin{equation}\label{eq:born}
\langle|\int d^3{\bf r} e^{2\pi i n_0 x}\,\delta\rho({\bf r})|^2\rangle = [\chi_0^\infty(2\pi n_0)e^2/\pi n_0^2]^2 N_i,
\end{equation}
where $N_i$ is the total number of impurities in the sample. The right hand side of Eq. (\ref{eq:born}) is $\sim S_0^8$ and therefore the QPS rates are strongly suppressed in wires, whose diameter significantly exceeds the Fermi wavelength.

Then we obtain an estimate for the average frequency:
\begin{equation}\label{eq:final}
\langle\omega_0\rangle \sim  {\alpha_s^2 c_s N_i^{1/2}\over \xi_s ^2 a_B \rho_0^4 S_0^5}\times\left({\xi_s\over L}\right)^{\alpha_s},
\end{equation}
where $a_B$ is Bohr radius. Let us estimate the value of $\langle\omega_0\rangle$ for typical experimental parameters.
We consider a $MoGe$ wire, e.g. Ref. \cite{bez}, of diameter $2r_0 = 5\, nm$ and length $L=1\, \mu m$.  The density of conduction electrons $\rho_0$ can be estimated from the data on conductivity in the normal state, e.g. Ref. \cite{beasley}, $\sigma = e^2 \rho_0 /k_F l$. For an amorphous material, such as $MoGe$, the mean free path is of the order of interatomic distance $l\sim 4 \AA$, e.g. Ref. \cite{beasley}, and thus we obtain $\rho_0\simeq 10^{28}\, m^{-3}$. In order to estimate $\alpha_s$ and $c_s$ we need values of $\rho_s$ and $C$. Assuming that the wire is covered by a dielectric material, e.g. Ref. \cite{bez,atom,exp}, we estimate that $C\sim 1$. The density of superconducting electrons is related to the normal electron density as $\rho_s \simeq \rho_0 (l/\xi_s)^2$ \cite{AGD}. Taking $\xi_s=8\, nm$, e.g. Refs. \cite{bez, beasley}, we obtain $\rho_s\simeq 10^{25} m^{-3}$ and $\alpha_s\simeq 0.5$ and $c_s\simeq 7\times 10^5\,  m/s$. Finally, the number of impurities $N_i$ is of the order of the total number of atoms in the wire and so we take $N_i\simeq 3\times 10^5$. With such parameters we obtain $\langle\omega_0\rangle\sim 10^3 \, s^{-1}$. As we pointed out above, the QPS rate is significantly suppressed by the small ``scattering amplitude'', e.g. Eq. (\ref{eq:born}), due to the high value of the QPS momentum. The value of $\langle\omega_0\rangle$ can greatly increase in thinner wires, e.g. Eq. (\ref{eq:final}): For wires with diameter $2\, nm$ we obtain $\langle\omega_0\rangle\sim 10^7 \, s^{-1}$. The coherent QPS in such wires can presumably be detected by inductively coupling the ring to a SQUID.

In summary we have studied a possibility for generation of coherent QPS in small superconducting rings. We have evaluated the frequency of such transitions and found that it is appreciable only in ultrathin wires, i.e., with diameters not too strongly exceeding the electron's Fermi wavelength. We have also shown that such frequency is a non-self-averaging, sample-dependent quantity.

{\acknowledgements We thank M. Boshier, I. Martin, S. D. Snyder and E. Timmermans for valuable discussions and comments. The work is supported by the US DOE.}

\end{document}